\documentclass{iopart}

\usepackage{graphicx}
\begin{document}

\title[The effect of ion-implantation on the electrical transport
properties of Si-SiO$_{\textrm{2}}$ MOSFET's]{The effect of
low-energy ion-implantation on the electrical transport properties
of Si-SiO$_{\textrm{2}}$ MOSFETs}

\author{D R McCamey$^{1,2}$, M Francis$^{2}$, J C McCallum$^{1,3}$,
\\ A R Hamilton$^{1,2}$, A D Greentree$^{1,3}$ and R G  Clark$^{1,2}$ }

\address{$^{1}$ Australian Research Council Centre of Excellence for Quantum Computer Technology}
\address{$^{2}$ School of Physics, University of New South Wales, NSW 2052, Australia}
\address{$^{3}$ School of Physics, University of Melbourne, VIC 3010, Australia}

\ead{dane.mccamey@unsw.edu.au}

\begin{abstract}

Using silicon MOSFETs with thin (5nm) thermally grown SiO$_{2}$
gate dielectrics, we characterize the density of electrically
active traps at low-temperature after 16keV phosphorus
ion-implantation through the oxide. We find that, after rapid
thermal annealing at 1000$^{\textrm{o}}$C for 5 seconds, each
implanted P ion contributes an additional $0.08 \pm 0.03$
electrically active traps, whilst no increase in the number of
traps is seen for comparable silicon implants. This result shows
that the additional traps are ionized P donors, and not damage due
to the implantation process. We also find, using the room
temperature threshold voltage shift, that the electrical
activation of donors at an implant density of $2\times10^{12}
\textrm{cm}^{-2}$ is $\sim100\%$.

\end{abstract}


\section{Introduction}

Device fabrication involving ion implantation is widespread in the
semiconductor industry, with applications ranging from ohmic
contacts to shallow junctions \cite{sze,shallow}. Methods to
increase the performance of classical transformers \cite{Shinada},
new types of classical computation \cite{cole}, and a number of
solid state implementations of quantum computation (QC)
\cite{Kane,McKinnon,Schenk,Vrijen,Hollenberg,Clark}, all involving
ion-implantation of few or single ions, have been proposed. In the
proposal by Hollenberg \emph{et al} \cite{Hollenberg} for example,
the position of the donor electron on one of two implanted
\cite{TopDown} closely spaced P donors in Si act as the qubit.

In all of these devices, electrically active defects caused by ion
implantation must be eliminated so that operations involving
single electrons are not compromised. To ensure that this
requirement is met, it is important to characterize the effects of
ion implantation on the trap density. It is also important that
all implanted donors be activated for the device to function
correctly. Given that these devices are intended to be operated in
the few or single electron regime, a non-activate donor anywhere
in the device would strongly inhibit device function.

Previous studies of damage caused by ion implantation have mainly
involved capacitive measurements (eg capacitance-voltage, Deep
Level Transient Spectroscopy \cite{Yamasaki}) or have focused on
room temperature measurements \cite{Peterstrom}. This paper
focuses on the characterization of implantation-induced defects in
MOSFET devices using DC transport measurements at low-temperature,
where silicon-based quantum computer devices are most likely to be
operated. We also study the donor activation of these implants.

\section{Method}

MOSFETs are convenient to fabricate, have widely understood
characteristics, and allow electrical measurements to be
restricted to near the Si-SiO$_{2}$ interface \cite{Sun}, the area
of interest for low-energy implantation studies. Additionally, the
processing used in the fabrication of MOSFETs is expected to be
compatible with the fabrication processes used, for example, in
silicon-based quantum computer devices. This allows the
possibility of fabricating on-chip characterization devices. For
these reasons, MOSFETs were used to characterize the implant
damage in this study.

\subsection{Mobility and Critical Density at 4.2K}\label{mobsect}

Numerous methods exist for characterizing the electrically active
damage, or traps, at room temperature \cite{Witc}. At cryogenic
temperatures, where standard capacitance based methods fail due to
the long thermal emission time, a number of more complex methods
also exist \cite{Divak}. Some simpler methods using the Hall
effect involve comparing measured values of carrier density with
theoretical predictions \cite{saks,fowler}. We use a method that
involves only conductance and Hall measurements, removing errors
associated with comparison to theory and allowing for ease of
measurement.

At low temperatures, the mobility, $\mu$, of an electron in the
inversion layer of a MOSFET as a function of the carrier density,
$n$, is characterized by a critical density, $n_{crit}$, below
which no conduction occurs. The lack of conductivity below the
critical density is due to a freeze out of free carriers due to
impurity binding \cite{Dassarma}. The method used to characterize
the trap density in this work is based on the above property. The
mobility as a function of the carrier density was determined from
Hall measurements. The critical density was then determined by
extrapolating the linear region of $\mu$ vs $\log(n)$ above the
critical density to zero. The trap density of the device is taken
to be the critical density.

Measurements at 4.2K were performed to determine the number of
electrically active traps present at these temperatures. The
4-terminal resistivity, $\rho$, of the inversion layer was
measured at numerous gate voltages. Hall measurements were taken
from B=0 to 0.5T, for at least 5 gate voltages per sample. The
carrier density as a function of gate voltage was determined from
these measurements. The mobility was determined using both the
resistivity and the carrier density by the relation \cite{sze}

\begin{equation}
\mu = 1/(\rho n e).
\end{equation}

\subsection{Threshold Voltage Shift at Room Temperature}

To determine the fraction of implanted ions that were activated,
the threshold voltage shift was analyzed. Assuming the ions are
implanted to a constant depth (ie a delta function approximation),
the relation between doping and threshold shift is given by
\cite{sze}

\begin{equation}
n_{\textrm{act}} = \frac{\Delta
V_{\textrm{th}}C_{\textrm{imp}}}{q}
\end{equation}

where $n_{\textrm{act}}$ is the number of activated donors per
unit area, $\Delta V_{\textrm{th}}$ is the change in threshold
voltage, and $C_{\textrm{imp}}$ the capacitance per unit area
between the implanted ion layer and the gate. As the implant
distribution sits 20nm into the silicon, the capacitance is the
sum of that due to the oxide and the silicon layer, ie

\begin{equation}
\fl C_{\textrm{imp}} = [(1/C_{\textrm{ox}}) +
(1/C_{\textrm{Si}})]^{-1}=[(d_{\textrm{ox}}/\epsilon_{\textrm{ox}}
\epsilon_{0})+(d_{\textrm{Si}}/\epsilon_{\textrm{Si}}\epsilon_{0})]^{-1}
\end{equation}

where $d_{\textrm{ox}}$ is 5nm, $\epsilon_{\textrm{ox}}$ is 3.7,
$d_{\textrm{Si}}$ is 20nm and $\epsilon_{\textrm{Si}}$ is 11.9.

It is important to note that this method is only valid for ionised
impurities, and for this reason the measurements were performed at
room temperature. The effect of incomplete ionisation
\cite{ionisation} is discussed with the results.

To confirm the validity of the delta approximation, once the
number of ionised impurities was determined, the expected
threshold shift was calculated using a model for the implants that
consisted of a series of very thin, uniform implant regions that
very closely followed the actual implant distribution. The
threshold shift due to each of these was determined and the total
shift found. There was a negligible ($<1\%$) difference between
these methods, indicating that the original single delta
approximation was sufficient.

Modelling, using a one-dimensional Poisson solver
\cite{1Dpoisson}, of both implanted and unimplanted devices was
also performed. The carrier density as a function of depth for a
number of gate voltages was determined, and integrated to find the
carrier density as a function of gate voltage. The threshold in
these simulations was compared to the experimentally measured
threshold, again confirming the density of activated ions.

\subsection{Device Fabrication}

The devices used in this study, Hall-bar geometry MOSFETs, were
fabricated on a high resistivity ($>5000 \Omega$.cm) n-type Si
$<$100$>$ substrate. After etching in a 10\% HF solution for 10
seconds to remove the native oxide, a 5nm thermal oxide layer was
grown. Two implant species, P and Si, were used,  P as this is the
most common donor in silicon and has important applications in QC
proposals, and the Si as a control. A number of devices were then
implanted with either P at 16keV or Si at 15keV.  This results in
an implant distribution centered approximately 20nm into the
silicon, with a straggle of approximately 7nm \cite{sze}. A number
of different doses were implanted, ranging from no implant to
$5\times10^{12} \textrm{cm}^{-2}$. Rapid thermal annealing at
1000$^{\textrm{o}}$C for 5 seconds in a N$_{2}$ ambient was then
applied to all devices. For both the gate and ohmic metallization
200nm of Aluminium was used. Following metalisation, the devices
were annealed at 400$^{\textrm{o}}$C for 15 minutes in forming gas
(5\% Hydrogen, 95\% Nitrogen).

\section{Results}
\subsection{Implant Activation at Room Temperature}

\begin{figure}
\centering
  \includegraphics[width=8cm]{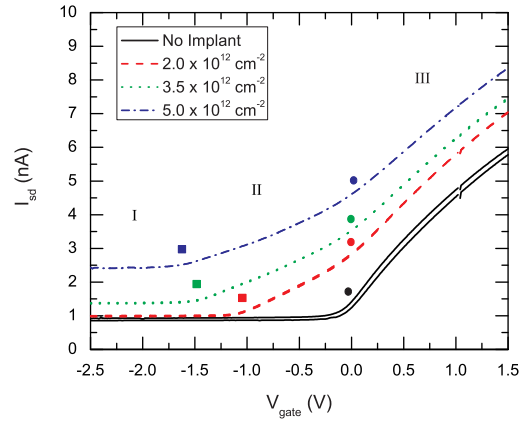}\\
  \caption{Source drain
current as a function of Gate voltage, showing
  threshold voltage shift as a function of implant density. Measurements
  taken with a 100$\mu$V source drain voltage at room temperature.
  The squares indicate the threshold voltage after
implantation. The circles show the location of the kink in the
current at a similar voltage to the threshold in the unimplanted
device.I, II, and III indicate three different conduction regimes,
which are discussed in the text.}\label{RTVth}
\end{figure}

\begin{figure}
\centering
  \includegraphics[width=8cm]{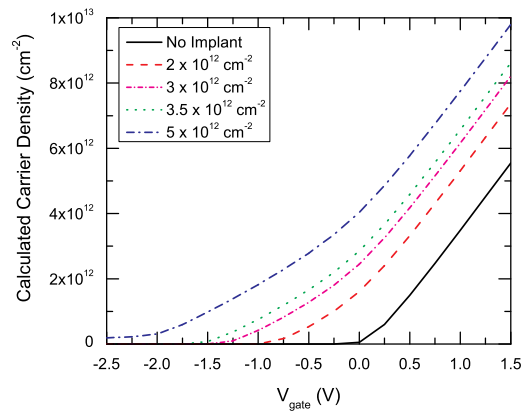}\\
  \caption{Calculated carrier density vs gate voltage, T=300K, implant
  range = 20nm, implant straggle = 7nm.}\label{calcRTnvsVg}
\end{figure}

\begin{figure}
\centering
  \includegraphics[width=8cm]{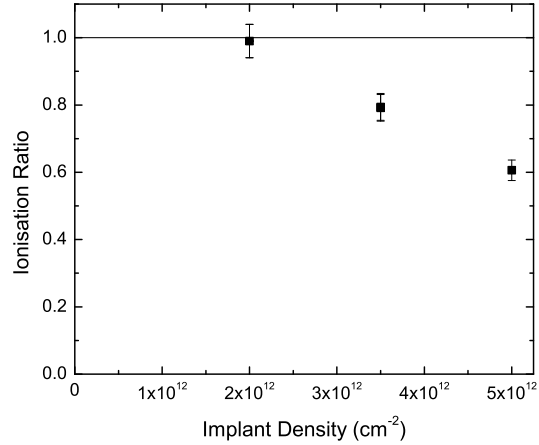}\\
  \caption{Measured ionisation ratio ( number of activated ions times probability of ionisation divided by number of implanted ions) as a function of implant density for P
  implants at 16keV after rapid thermal annealing.}\label{Activation}
\end{figure}

The I-V characteristics of the MOSFETs were measured at room
temperature. The results for different P implant densities are
shown in figure \ref{RTVth}. The threshold voltage was taken to be
the intersection of the linear fit to the sub-threshold current
and the linear fit to the current in the region just above the
point where the current deviates from the sub-threshold current.
As expected, the threshold voltage is decreased by the addition of
n-type dopants. The Si implanted control devices did not show a
decrease in threshold voltage, which shows that the change in
threshold is due to the incorporation of donors, and not the
creation of charged damage during implantation.

The activation ratio is defined as the number of activated P atoms
(ie those sitting in substitutional sites) divided by the number
of implanted P ions. The threshold voltage shift is determined by
the number of charged P donors, which is the product of the number
of activated donors and the probability that they are ionised.
Figure \ref{Activation} shows the measured ionisation ratio
(activation ratio times probability of activation) as a function
of the implant dose. The ionisation ratio is found to decrease
with increasing implant dose. This is expected, given that the
fraction of ionised donors is known to decrease with increasing
density of donors \cite{Singh}, due to the increased number of
dopant states available, even at room temperature
\cite{ionisation}.

We note that recent studies of P-implanted silicon by spreading
resistance analysis (SRA) have shown the opposite trend - that the
apparent activation ratio increases with increasing implant dose
\cite{Schenk,Park}. This discrepancy might possibly be explained
by the presence of a native (poor quality) oxide with a high trap
density in very close proximity to the implanted region, caused by
cutting the wafer at a very shallow angle. If a significant
density of electrons are caught in traps at the native oxide
interface they do not contribute to conduction, so that the number
of free electrons appears lower than that expected due to the
implant dose. This effect would be particularly significant at low
implant dose, where the native oxide trap density is much higher
than the P density, but would be less important at high doses,
leading to an apparent activation ratio that increases with
increasing implant dose, as observed in \cite{Schenk,Park}. If
this is the case, it means that spreading resistance measurements
are not well suited to measurements of low dose, near surface
implants - however, more work on understanding the difference
between these two measurement techniques is required.

It is significant however that after RTA and for an implant dose
of $2 \times 10^{12} \textrm{cm}^{-2}$  with an average spacing of
$\sim 7$nm, the ionisation ratio is $0.99\pm0.05$, indicating
almost complete activation and ionisation. Assuming that this near
complete activation holds for doses below $2 \times 10^{12}
\textrm{cm}^{-2}$, this result suggests that it will be possible
to fabricate a device where the donor spacing is $\sim 20$nm, such
as the qubit proposed by Kane \cite{Kane}, with almost complete
donor activation.

\begin{figure}[h]
\vspace{1cm}
\begin{minipage}[t]{8cm}
  \begin{center}
   Unimplanted
   \newline
   \includegraphics[width=7.9cm]{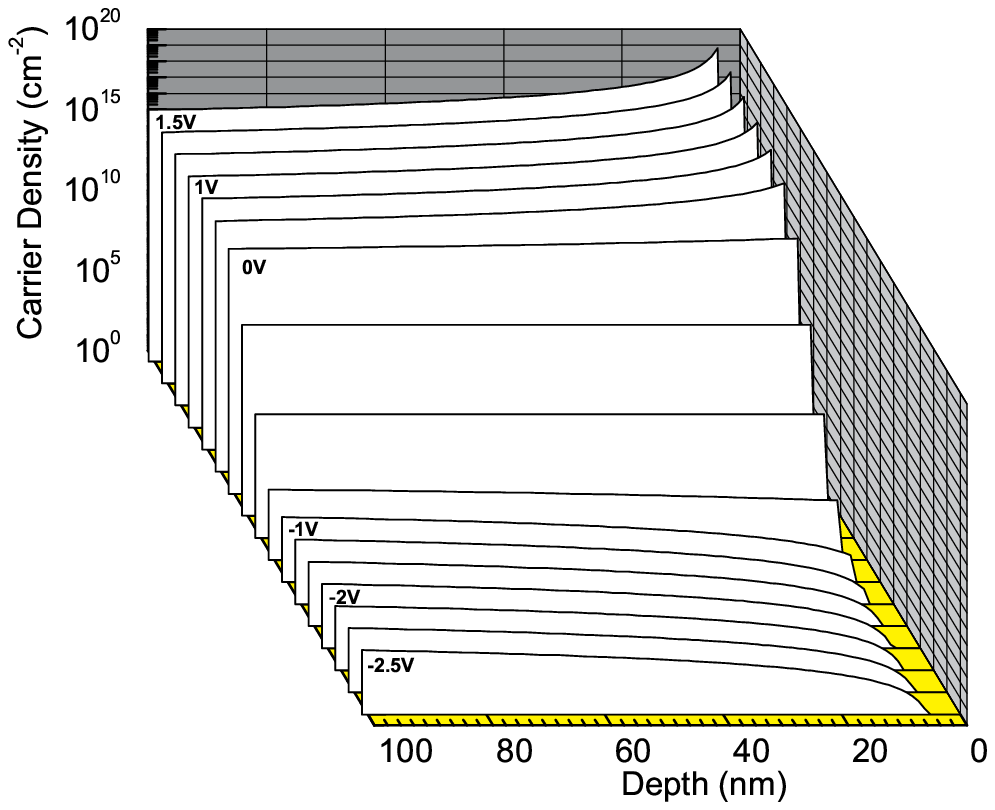}
  \end{center}
\end{minipage}
\hfill
\begin{minipage}[t]{8cm}
  \begin{center}
    Implanted
    \newline
    \includegraphics[width=7.9cm]{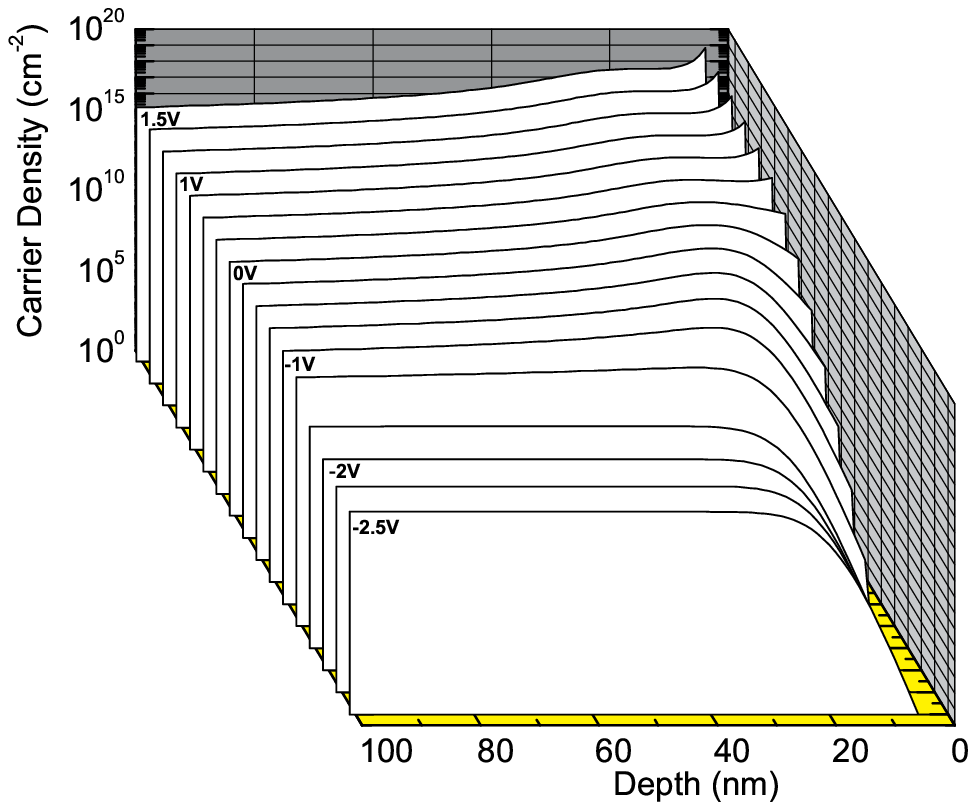}
  \end{center}
\end{minipage}
\caption{\label{nvddimp3p5}Calculated carrier density as a
function of depth and Gate Voltage, at room temperature for an
unimplanted MOSFET (left) and for a MOSFET with a P implant dose
of $3.5 \times 10^{12}$ cm$^{-2}$ (right). T = 300K.}
\end{figure}

Another interesting feature to note is the appearance of three
distinct conduction regimes in the implanted devices, labeled I,
II and III in figure \ref{RTVth}. In region II, from threshold to
the threshold associated with the unimplanted device, the device
current increases near linearly. Above the unimplanted threshold,
when the device enters inversion, the device current returns to
the expected form but with an offset (region III). In region I,
below threshold, there is a constant current unaffected by the
gate voltage, but which increases with implant density. The source
of this current is unknown. A possible explanation is that the
silicon dioxide used as an implant mask was not thick enough,
resulting in an implant in the area outside of the gated area.
However, the thickness of the implant mask ($\sim200$nm) should be
more than adequate to stop all of the implanted ions, which have a
range of $\sim15$nm in SiO$_{2}$ at 15keV \cite{sze}.
Alternatively, Poisson modelling shows that there is a small
increase in the carrier density deep in the silicon
($\sim50$nm)for implanted devices at negative gate voltages (see
figure \ref{nvddimp3p5}). This would explain the increased
conduction, except that the carrier density at negative voltages
is 5 orders of magnitude smaller than that at positive voltages,
which does not explain the high current observed. More work is
needed to understand the cause of this conduction.

Modelling of the devices using a one-dimensional Poisson solver
\cite{1Dpoisson} was undertaken, and the carrier density as a
function of depth for a number of gate voltages are presented in
figure \ref{nvddimp3p5}. In the implanted devices, an increase in
the carrier density, centered at the mean implant depth,
$d_{\textrm{Si}}$, is observed. The density in this region
increases with gate voltage, until the unimplanted threshold is
reached. For higher gate voltages, the carrier density is
dominated by the inversion layer adjacent to the
Si-SiO$_\textrm{2}$ interface.

The total carrier density at a given gate voltage and implant
density was determined by integrating the calculated carrier
density to a depth of 100nm. The results are shown in figure
\ref{calcRTnvsVg}. These traces have a similar form to the
measured current (figure \ref{RTVth}) showing the two distinct
conduction regimes above threshold, but not the sub-threshold
current. The effect of scattering can be seen in the deviation of
the measured current from the calculated carrier density. The
threshold however should not be effected by the scattering
mechanisms, and the current should go to zero at the same gate
voltage as the carrier density.

Calculations were performed for devices with P implants of 2, 2.7
and 3$\times 10^{12}$ cm$^{-2}$, the density of activated donors
determined from the threshold voltage shift. The threshold found
in these calculations was in good agreement with that observed
experimentally, and better than those simulations where the
incomplete ionisation of donors was not considered. This indicates
that the analysis of the activation ratio is correct.

To summarize, we find that there is a decrease in the threshold
voltage due to the implantation of donors, although the magnitude
of the decrease is smaller than expected. This may be due to the
incomplete ionisation of the implanted donors, which serves to
limit their effect on the threshold voltage. We find that for an
implant of 2$\times 10^{12}$ cm$^{-2}$, the activation is near
100\%.

\subsection{Damage due to Implantation}
\begin{figure}[h]
\vspace{1cm}
\begin{minipage}[t]{7.5cm}
  \includegraphics[width=7cm]{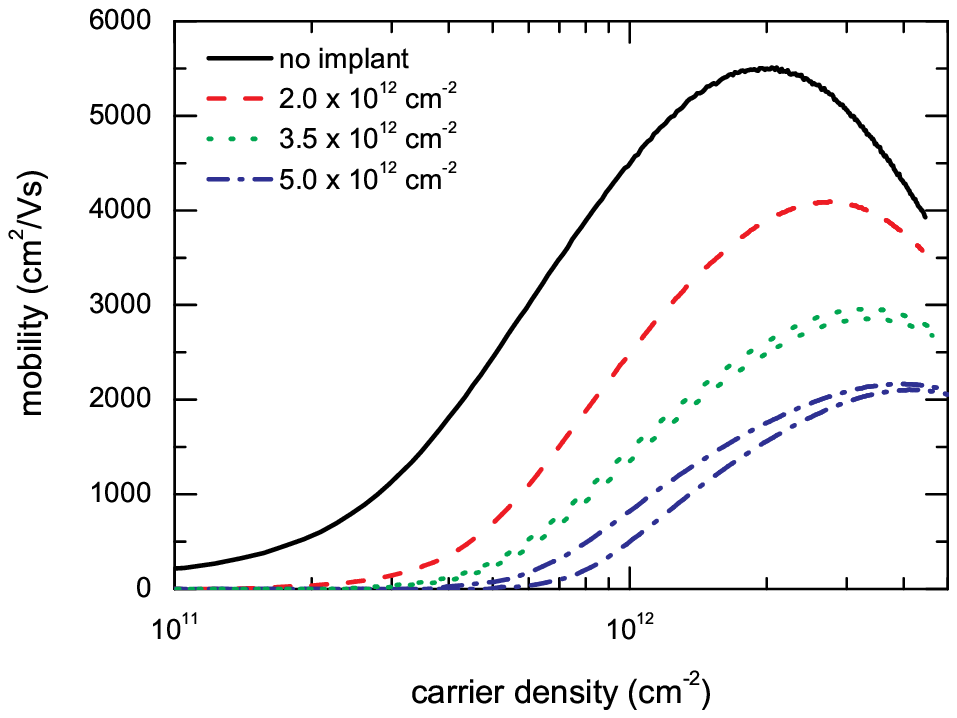}
\end{minipage}
\hfill
\begin{minipage}[t]{7.5cm}
  \includegraphics[width=7cm]{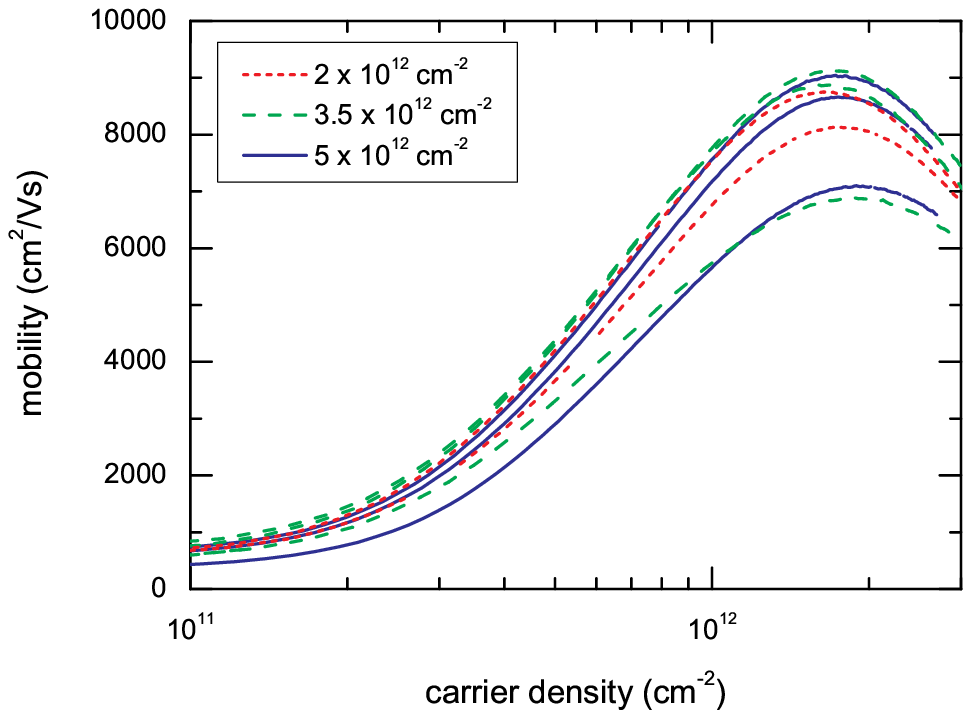}
\end{minipage}
\caption{\label{Pmob}\label{Simob}Mobility as a function of
carrier density for MOSFET's with various
  P (left) and Si (right) implantation densities.}
\end{figure}

We now turn to consider the damage caused by the implantation
process. Figure \ref{Pmob} shows the mobility of P implanted
MOSFET's, at a number of different doses. The traces show the
usual form of the low temperature mobility, that is, at low
density, an increasing mobility limited by impurity scattering,
and at high density, a decreasing mobility limited by scattering
due to interface roughness. Traces for a number of different
implant densities are shown.

\begin{figure}[h]
\vspace{1cm}
\begin{minipage}[t]{7.5cm}
  \includegraphics[width=7cm]{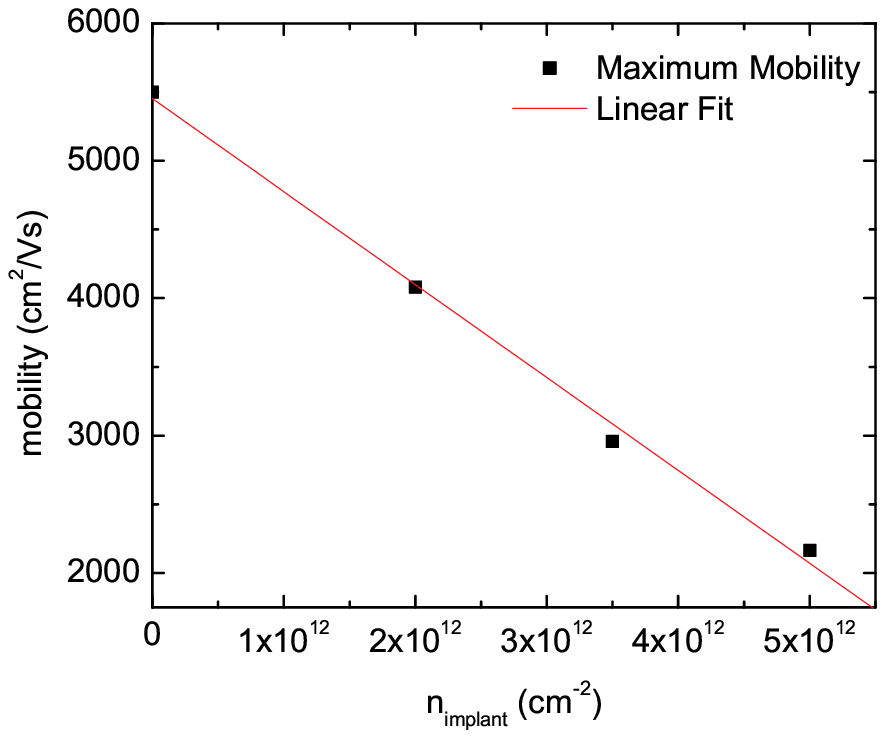}
\end{minipage}
\hfill
\begin{minipage}[t]{7.5cm}
  \includegraphics[width=7cm]{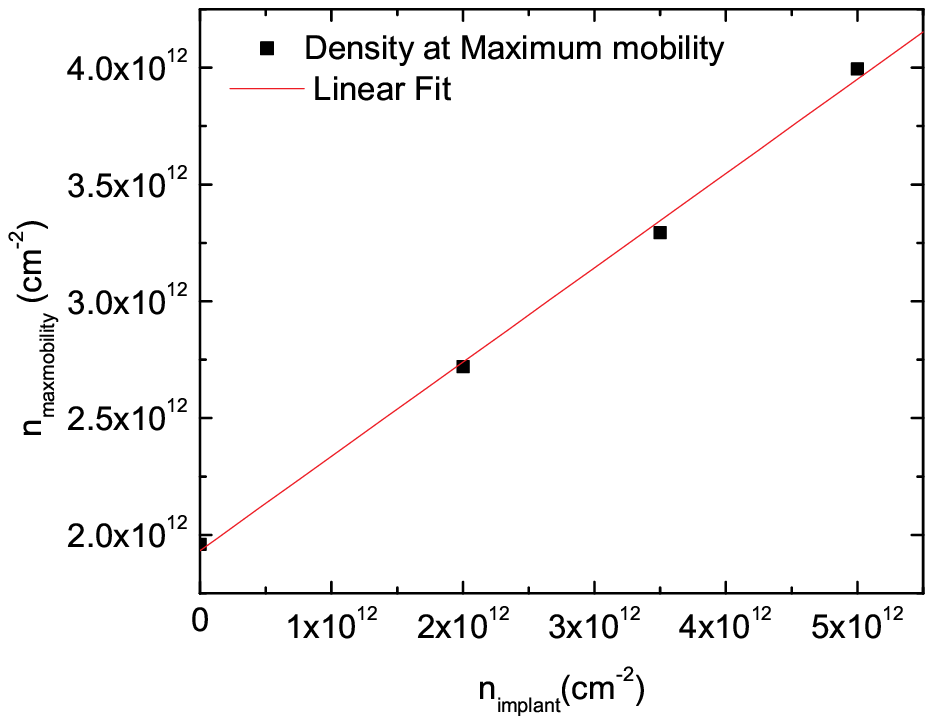}
\end{minipage}
\caption{\label{maxmobvsn}\label{maxmobnvsn}Maximum mobility
(left) and carrier density at which maximum mobility occurs
(right) as a function of P implant density. Both the decrease in
mobility and the movement of peak mobility to higher density
indicate an increase in charged impurity scattering as the implant
dose is increased.}
\end{figure}

In the P implanted devices, the maximum mobility is seen to
decrease and move to higher carrier densities (figure
\ref{maxmobnvsn} ) as the implant dose is increased, indicating an
increase in ionized impurity scattering. For Si implanted devices
the mobility, shown in figure \ref{Simob}, does not show this
effect. As the Si and P ions are of similar mass, and are
implanted at similar energies, they should cause similar damage to
the Si lattice during implantation. As the decrease in mobility is
not seen in the Si implanted devices, the cause of the increased
scattering does not appear to be the implantation process. This
result suggests that the RTA is able to repair all damage from
implantation. An alternate explanation is that stable P-related
defects may form during the annealing process accounting for the
increased trap density, but we believe that this will be a
secondary effect when compared to the lattice damage caused during
implantation.

It is important to note that there are variations in the
mobilities across wafers and from chip to chip, however, the
deviation between devices on a similar chip is much smaller than
that from chip to chip from the same wafer, which is smaller again
than that from wafer to wafer. To combat this, the measurements
for a given implant species were taken from single wafers. The
wafers were cleaved into quarters, with each quarter implanted at
a different dose, and one quarter being left unimplanted. As a
result, it is difficult to compare numerically the results from
wafer to wafer. This explains the higher peak mobility in the Si
implanted devices compared to the P implanted devices.

\begin{figure}
\centering
  \includegraphics[width=8cm]{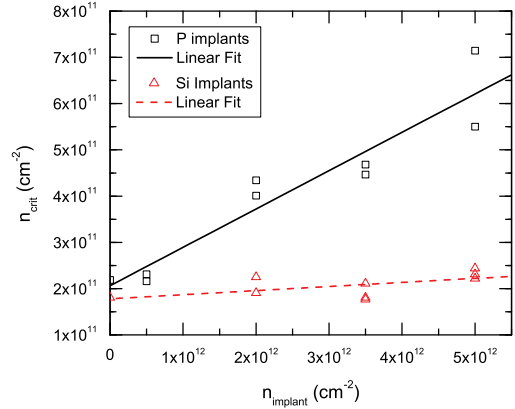}\\
  \caption{Trap density vs Implant density for both P and Si implanted devices.
  Linear fits to the data are shown. The data for P implants at
  $5 \times 10^{11} \textrm{cm}^{-2}$ is for a device fabricated on a separate wafer. }\label{TDvsID}
\end{figure}

The critical density of both P and Si implanted devices is shown
in figure \ref{TDvsID}. For P implanted devices, the critical
density increases linearly with implant density. Equating the
critical density with the trap density \cite{Dassarma} shows that
each implanted ion has an equivalent effect to the creation of
$0.08 \pm 0.03$ additional traps. For Si implanted devices, this
effect decreases to $0.009 \pm 0.005$ additional traps.

The above results can be explained by the straggle of the
implanted ions. Modelling of the implanted ion distribution using
data from \cite{sze} shows that, for the implant energy used,
approximately 10\% of implanted ions sit within 10nm of the
interface. Modelling of the wavefunction using \cite{1Dpoisson}
and other studies \cite{stern} shows that electrons in the
inversion layer are localized to within approximately 10nm of the
Si-SiO$_\textrm{2}$ interface. Hence, only about 10\% of the
implanted ions will interact strongly with the electrons in the
inversion layer.

Below threshold, there are no electrons in the 2DEG, and all donor
electrons are bound to the P implants. Above threshold, when the
channel is populated with electrons, the additional electrons
serve to screen the P donors, and they are unable to localize any
electron \cite{stern}. At this point, the ionized P donor serves
to scatter the electrons in the 2DEG, leading to the increased
ionized impurity scattering seen in figure \ref{Pmob}. This also
explains the movement of the critical density, as the electrons
that are localized below the threshold are free to contribute to
the Hall voltage, and thus the carrier density, above threshold.

The increase in the critical density (8\% of implant) is in good
agreement with the number of implanted donors near the 2DEG
($\sim100\%$ of implant), again suggesting that RTA is able to
remove all damage caused by implantation, and that the increased
scattering and trapping is due only to the P donors close to the
Si-SiO$_{2}$ interface.

Another important characteristic is the density of electron traps
in the unimplanted device. This was found to be $2.1 (\pm0.3)
\times 10^{11} \textrm{cm}^{-2}$. This equates to a trap spacing
of $21.8 \pm 1.7$ nm. For a Kane architecture quantum computer,
this is approximately the qubit-qubit spacing (20nm). This means
that for every implanted P donor there is approximately one
electron trap, which will interfere with device operation, either
by localizing the donor electron or by other methods. While this
density of traps may be suitable for fabrication of a small number
of silicon-based qubits, it will need  to be reduced, for example
by improving the oxide quality, for the large scale (many qubit)
implementations that are proposed.

\section{Conclusions}

We have characterizing the density of electrically active traps at
low temperature, using DC conductance measurements in silicon
MOSFETs. We determine the effect of low-energy low-density Si and
P ion implantation into a Si-SiO$_{2}$ system, and found that the
Si implantation had no effect on the low temperature trap density,
whilst the P implantation resulted in an additional
$0.08(\pm0.01)$ traps per implanted ion. We interpret these
additional traps to be P donors in the conduction channel acting
to localize donor electrons, and not damage due to implantation.

We also find that the electrical activation of the implanted
donors to be near complete at a dose of $2 \times
10^{12}\textrm{cm}^{-2}$, falling to approximately 60\% at a dose
of $5 \times 10^{12}\textrm{cm}^{-2}$. This indicates that a
device that requires a donor spacing of approximately 20nm, such
as the Kane quantum computer, can be fabricated with near 100\%
activation using ion implantation.

\ack The authors thank E. Gauja and S. Angus for assistance in
device fabrication, R. P. Starrett and D. Barber for assistance
with construction of low temperature measurement equipment and A.
S. Dzurak for helpful discussions. This work was supported in part
by the Australian Research Council, the Australian Government, the
U.S. National Security Agency, the Advanced Research and
Development Agency, and the U.S. Army Research Office under
Contract No. DAAD19-01-1-0653.


\section*{References}

\begin{thebibliography}{99}

\bibitem{sze} Sze S M 1975 {\it Physics of Semiconductor Devices
2nd Edition} (New York: John Wiley \& Sons )

\bibitem{shallow} Jones E C and Ishida E 1998 Shallow junction
doping technologies for ULSI {\it Materials Science and
Engineering: R: Reports} {\bf R24} 1

\bibitem{Shinada} Shinada T, Koyama H, Hinoshita C, Imamura K and
Ohdomari I 2002 Improvement of focused ion-beam optics in
single-ion implantation for higher aiming precision of one-by-one
doping of impurity atoms into nano-scale semiconductor devices
{\it Japanese Journal of Applied Physics Part 2 - Letters} {\bf
41} L287

\bibitem{cole} Cole J H, Greentree A D, Wellard C J, Hollenberg L C L
and Prawer S 2004 Quantum-Dot Cellular Automata using Buried Dopants
{\it eprint arXiv:cond-mat/0407658}

\bibitem{Kane} Kane B E 1998 A silicon-based nuclear spin quantum
computer {\it Nature} {\bf 393} 133

\bibitem{McKinnon} McKinnon R P, Stanley F E,
Lumpkin N E, Gauja E, Macks L D, Mitic M, Chan V, Peceros K,
Buehler T M, Dzurak A S, Clark R G, Yang C, Jamieson D N and
Prawer S D 2002 Nanofabrication processes for single-ion
implantation of silicon quantum computer devices {\it Smart
Materials and Structures} {\bf 11} 735

\bibitem{Schenk} Schenkel T, Persaud A, Park S J, Nilsson J,
Bokor J, Liddle J A, Keller R, Schneider D H, Cheng D W and
Humphries D E 2003 Solid state quantum computer development in
silicon with single ion implantation {\it Journal of Applied
Physics}{\bf 94} 7017

\bibitem{Vrijen} Vrijen R, Yablonovitch E, Wang K, Jiang H W,
Balandin A, Roychowdhury V, Mor T and DiVincenzo D 2000
Electron-spin-resonance transistors for quantum computing in
silicon-germanium heterostructures {\it Physical Review A} {\bf
62} 012306

\bibitem{Hollenberg} Hollenberg L C L, Dzurak A S, Wellard C, Hamilton A R,
Reilly D J, Milburn G J and Clark R G 2004 Charge-based quantum
computing using single donors in semiconductors
 {\it Physical Review B} {\bf 69} 113301

\bibitem{Clark} Clark R G, Brenner R, Buehler T M, Chan V, Curson N J, Dzurak A S,
Gauja E, Goan H S, Greentree A D, Hallam T, Hamilton A R,
Hollenberg L C L, Jamieson D N, McCallum J C, Milburn G J, O'Brien
J L, Oberbeck L, Pakes C I, Prawer S D, Reilly D J, Ruess F J,
Schofield S R, Simmons M Y, Stanley F E, Starrett R P, Wellard C
and Yang C 2003 Progress in silicon-based quantum computing {\it
Philosophical Transactions of the Royal Society of London Series A
- Mathematical Physical and Engineering Sciences} {\bf 361} 1451

\bibitem{TopDown} Dzurak A S, Hollenberg L C L, Jamieson D
N, Stanley F E, Yang C, B\"{u}hler T M, Chan V, Reilly D J,
Wellard C, Hamilton A R, Pakes C I, Ferguson A G, Gauja E, Prawer
S, Milburn G J and Clark R G 2003 Charge-based silicon quantum
computer architectures using controlled single-ion implantation
{\it eprint arXiv:cond-mat/0306265}

\bibitem{Yamasaki} Yamasaki K, Yoshida M, Sugano T 1979 Deep level
transient spectroscopy of bulk traps and interface states in Si
MOS diodes {\it Japanese Journal of Applied Physics} {\bf 18} 113

\bibitem{Peterstrom} Peterstr\"{o}m S 1993 Si-SiO$_{2}$ interface
trap density in boron and phosphorus-implanted silicon {\it
Applied Physics Letters} {\bf 63} 672

\bibitem{Sun} Sun S C and Plummer J D 1980 Electron mobility in
inversion and accumulation layers on thermally oxidized silicon
surfaces {\it IEEE Transactions on Electron Devices} {\bf 27} 1497

\bibitem{Witc} Witczak S C, Suehle J S and Gaitan M 1992 An
experimental comparison of measurement techniques to extract
Si-SiO$_{2}$ interface trap density {\it Solid-State Electronics
}{\bf 35} 345

\bibitem{Divak} Divakaruni R and Viswanathan C R 1995
Quasi-static behaviour of MOS devices in the freeze-out regime
1995 {\it IEEE Transactions on Electron Devices} {\bf 42} 87

\bibitem{saks} Saks N S, Ancona M G and Rendell R W 2002 Using
the Hall effect to measure interface trap densities in silicon
carbide and silicon metal-oxide-semiconductor devices.{\it Applied
Physics Letters} {\bf 80} 3219

\bibitem{fowler} Fang F F and Fowler A B 1968 Transport
Properties of Electrons in Inverted Silicon Surfaces. {\it
Physical Review} {\bf 169} 619

\bibitem{Dassarma} Das Sarma S and Hwang E H 1999 Charged
impurity-scattering-limited low-temperature resistivity of
low-density silicon inversion layers {\it Physical Review Letters}
{\bf 83} 164

\bibitem{ionisation} Xiao G , Lee J, Liou J J amd Ortiz-Conde A
1999 Incomplete ionization in a semiconductor and its implications
to device modelling {\it Microelectronics Reliability} {\bf 39}
1299

\bibitem{1Dpoisson} Snider G 1996 1D Poisson/Schrodinger
Solver \verb"http://www.nd.edu/~gsnider/"

\bibitem{Park} Park S -J, Persaud A,
Liddle J A, Nilsson J, Bokor J, Schneider D H, Rangelow I W,
Schenkel T 2004 Processing issues in top–down approaches to
quantum computer development in silicon {\it Microelectronic
Engineering} {\bf 73-74} 695

\bibitem{Singh} Singh J 2001 {\it Semiconductor Devices: Basic
Principles} (New York: John Wiley \& Sons ). In equation 2.34 on
page 80 we see that $\frac{n_{d}}{n+n_{d}} =
\frac{1}{\frac{N_{c}}{2N_{d}} \exp
\left[-\frac{(E_{c}-E_{d})}{k_{B}T}\right]+1}$.

\bibitem{stern} Stern F and Howard W E 1967 Properties of
semiconductor surface inversion layers in the electric quantum
limit {\it Physical Review} {\bf 163} 816

\end{thebibliography}
\end{document}